\documentclass{article}
\usepackage{ijcai25}

\usepackage{times}
\usepackage{soul}
\usepackage{url}
\usepackage[hidelinks]{hyperref}
\usepackage[utf8]{inputenc}
\usepackage[small]{caption}
\usepackage{graphicx}
\usepackage{amsmath}
\usepackage{amsthm}
\usepackage{amssymb}
\usepackage{booktabs}

\usepackage{algorithm}

\usepackage{algpseudocode}
\usepackage[switch]{lineno}

\usepackage{pifont} 

\usepackage{tikz}
\usepackage{array}

\usepackage{balance}

\title{NATLM: Detecting Defects in NFT Smart Contracts Leveraging LLM}
\author{
Yuanzheng Niu
\and
Xiaoqi Li\and
Wenkai Li
\\
Hainan University, Haikou, China\\
\emails
csxqli@ieee.org
}

\begin{document}

\maketitle

\begin{abstract}
Security issues are becoming increasingly significant with the rapid evolution of Non-fungible Tokens (NFTs). As NFTs are traded as digital assets, they have emerged as prime targets for cyber attackers. In the development of NFT smart contracts, there may exist undiscovered defects that could lead to substantial financial losses if exploited. To tackle this issue, this paper presents a framework called NATLM(\textbf{N}FT \textbf{A}ssistan\textbf{t} L\textbf{LM}), designed to detect potential defects in NFT smart contracts. The framework effectively identifies four common types of vulnerabilities in NFT smart contracts: ERC-721 Reentrancy, Public Burn, Risky Mutable Proxy, and Unlimited Minting.
Relying exclusively on large language models (LLMs) for defect detection can lead to a high false-positive rate. To enhance detection performance, NATLM integrates static analysis with LLMs, specifically Gemini Pro 1.5. Initially, NATLM employs static analysis to extract structural, syntactic, and execution flow information from the code, represented through Abstract Syntax Trees (AST) and Control Flow Graphs (CFG). These extracted features are then combined with vectors of known defect examples to create a matrix for input into the knowledge base. Subsequently, the feature vectors and code vectors of the analyzed contract are compared with the contents of the knowledge base. Finally, the LLM performs deep semantic analysis to enhance detection capabilities, providing a more comprehensive and accurate identification of potential security issues.
Experimental results indicate that NATLM analyzed 8,672 collected NFT smart contracts, achieving an overall precision of 87.72\%, a recall of 89.58\%, and an F1 score of 88.94\%. The results outperform other baseline experiments, successfully identifying four common types of defects.
\end{abstract}
 \vspace{-6pt}
\section{Introduction}



Ethereum \cite{2,gao2020checking,huang2021hunting,ma2024combining} introduced the paradigm of Turing-complete smart contracts \cite{3,li2017discovering,szabo1996smart,ante2022non,mikolov2013efficient}, which are self-executing agreements composed of code stored on the blockchain. The Ethereum Improvement Proposal EIP-721 \cite{5} introduced the ERC-721 token standard, enabling developers to create unique digital assets. Non-fungible tokens (NFTs) \cite{4} are non-replicable digital assets or unique identifiers managed on the blockchain, used to allocate, link, or prove ownership of distinct physical and digital goods. 
However, as the NFT market rapidly grows, the security issues surrounding smart contracts are gradually coming to light, becoming a key factor hindering the market's healthy development~\cite{niu2024unveiling,choromanski2020rethinking,beltagy2020longformer,srinivasa2020fast}. Once deployed, smart contracts are immutable~\cite{li2021hybrid}. While this immutability ensures transparency and reliability in transactions, it also means that attackers can exploit these flaws for malicious purposes if vulnerabilities exist, potentially leading to significant financial losses~\cite{zhong2023tackling}. These vulnerabilities mainly stem from logical flaws, contract design defects, and failure to consider potential security risks during development adequately~\cite{zou2025malicious,li2024detecting,chen2018system,liu2025sok}.

Traditional methods for detecting vulnerabilities in smart contracts, such as static analysis tools like Slither\cite{37slither} and Truffle\cite{truffle}, offer comprehensive reviews of smart contracts without executing the code, identifying common security flaws. However, these methods have certain limitations. Static analysis often relies on predefined rules and patterns, making it prone to false positives and false negatives when handling complex smart contracts, thus complicating the accurate identification of real security threats. 
The emergence of large language models (LLMs) \cite{llm2022emergent,malkov2018efficient,ain2019systematic,liang2024identity} has sparked widespread interest and research in academia. LLMs, such as CodeBERT \cite{2020codebert} and Gemini \cite{google_gemini_2024}, are models based on the Transformer \cite{transformers} architecture, possessing powerful natural language processing capabilities that enable them to learn complex semantics and contextual relationships from vast amounts of textual data. These models can not only understand and generate human-like natural language but can also be applied to code understanding and vulnerability detection. By learning from a large corpus of smart contract code and security audit reports, LLMs can effectively capture semantic features within the code, identify potential security risks, and generate detailed explanations of vulnerabilities. This capability gives LLMs a unique advantage in vulnerability detection, and they have already been employed in smart contract vulnerability detection \cite{2024smartcontractvulnhunt,zhu2022bytecode,zhang2024combining,pasqua2023enhancing} and software vulnerability detection \cite{2023software}  \cite{2024llbezpeky,bojanowski2017enriching,zhang2024acfix}.
However, LLMs primarily analyze code based on textual semantics but face limitations in precisely parsing code hierarchy, variable dependencies, and control structures. Additionally, LLMs may produce misleading results or interpretations, especially when dealing with complex or ambiguous code snippets, increasing the risk of false positives or negatives.
Current research finds \cite{david2023you} that while large language models (LLMs) can accurately identify some vulnerabilities in smart contract security audits, their high false positive rate still requires the involvement of manual auditors.

This study proposes a tool called NATLM to explore how combining large language models (LLMs) and static analysis methods can detect defects in NFT smart contracts. Specifically, we focus on four common NFT smart contract defects: ERC-721 Reentrancy, Risky Mutable Proxy, Public Burn, and Unlimited Minting defects \cite{yang2023definition}. The main contributions of this paper are as follows:
\begin{itemize}

\item To the best of our knowledge, this work is the first to propose NATLM, an innovative framework that combines static analysis with Gemini Pro 1.5 for detecting defects in NFT smart contracts. NATLM employs static analysis to extract critical features, including variables, functions, and control flow structures, directly from the contract’s codebase. These features constitute a comprehensive knowledge base, which enhances both the accuracy and the efficiency of defect detection.
\item Dual-Stage Detection with Feature Fusion for Improved Accuracy: NATLM employs a dual-stage detection strategy that combines feature fusion and advanced LLM analysis. 
\item We conduct extensive experiments to evaluate the effectiveness of NATLM across multiple
defects in NFT smart contracts. The overall precision is calculated to be 87.72\%.

\end{itemize}
 \vspace{-6pt}

\section{Related Work}
In the smart contract vulnerability detection field, numerous detection tools have emerged~\cite{li2024cobra,wu2025exploring,liu2024gastrace,bu2025enhancing,bu2025smartbugbert,li2024guardians,li2024defitail,li2024stateguard}.
The static analysis tool Slither\cite{37slither} detects potential vulnerabilities and coding errors by inspecting the smart contracts' bytecode and source code. Tools like Manticor e \cite{2019manticore} and Mythril\cite{Mythril2022survey} utilize symbolic execution to analyze smart contract behavior by exploring possible execution paths and identifying potential vulnerabilities and anomalies. Additionally, 
\cite{2021combining} combines graph neural networks with expert knowledge for smart contract vulnerability detection by transforming the control and data flow of the source code into a contract graph, normalizing it, and using temporal message propagation.

GPTLENS \cite{hu2023large} introduces an adversarial framework that divides the traditional single-stage detection process into two collaborative phases: generation and discrimination. In this framework, the LLM is designed to play two adversarial roles: the auditor and the critic. LLM4Fuzz \cite{shou2024llm4fuzz} integrates fuzz testing with LLMs to guide and organize fuzz testing activities. Liu et al. \cite{2024propertygpt} propose PropertyGPT, a system that leverages large language models like GPT-4 to generate formal verification properties for smart contracts automatically. The system addresses the challenges of developing properties that are compilable, appropriate, and runtime-verifiable by utilizing a vector database for property retrieval, iterative revision with external feedback, and formal verification through a dedicated prover.

 \vspace{-3.8pt}
\section{The NATLM Framework}
\begin{figure}[ht]
    \centering
    \includegraphics[height=0.22\textheight]{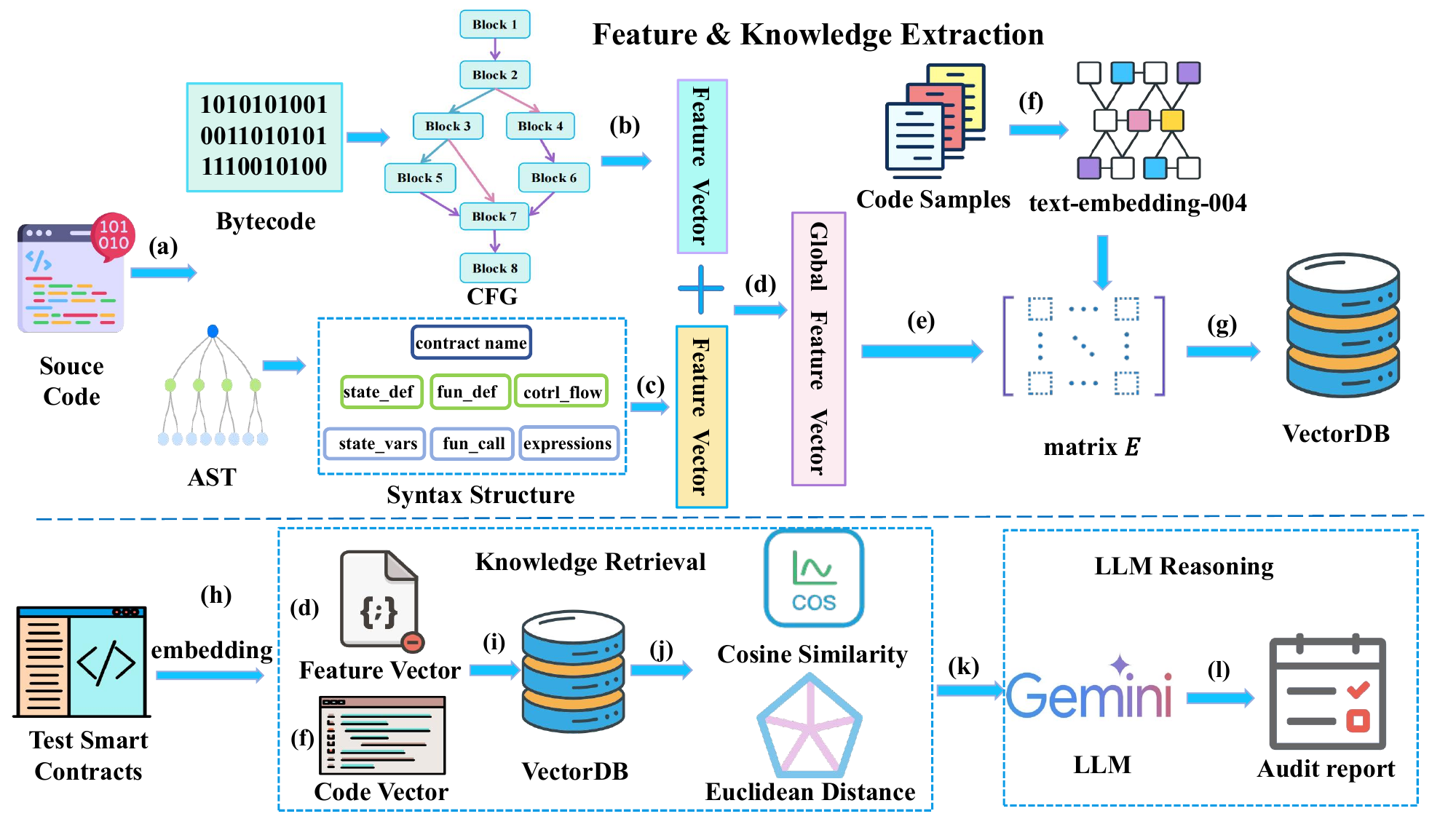}
    \caption{The Overall Framework of NATLM}
    \vspace{-6pt}
    \label{fig1l}
\end{figure}

In this section, we present the overall design of NATLM and provide a detailed explanation of its three phases from  Knowledge Base Building to  LLM Reasoning. The overall framework is illustrated in Figure \ref{fig1l}. The NATLM tool aims to detect defects in NFT smart contracts by combining static analysis with large language models (LLMs). The overall workflow of the tool consists of three main phases.
\vspace{-6pt}
\subsection{Feature \& Knowledge Base Extraction}
During the feature extraction and knowledge base construction phase, NATLM first parses the NFT smart contract source code to generate an Abstract Syntax Tree (AST), which reveals the structure and syntactical hierarchy of the code. We use the Solidity compiler (solc) to parse the NFT smart contract source code into an AST, mapping the structure of the code (such as function definitions, variable declarations, and control structures) into a tree format. The purpose of generating the AST is to capture the syntactic and semantic information within the contract, extracting key elements such as state variables, function definitions, and function call records while documenting the locations of these calls.

Once the AST is generated, we use CodeBERT\cite{2023smartcodebert} to extract feature vectors from the AST. CodeBERT is a pre-trained language model designed for code understanding, capable of capturing both syntactic and semantic information in the code. Before feeding the AST into CodeBERT, we first linearize the AST by performing a depth-first search (DFS) on the tree structure, flattening each node into a sequential order. 
The linearized sequence of nodes is then passed into CodeBERT's tokenizer, which splits complex identifiers into subtokens (e.g., `mintTokens` becomes `[mint, Tokens]`) to handle compound or rare terms. Special tokens `[CLS]` and `[SEP]` are added at the beginning and end of the sequence, respectively, to mark the sequence boundaries. Additionally, position embeddings are applied to each token to retain their relative order in the sequence.
The tokenized sequence, consisting of word embeddings and positional encodings, is passed through the embedding layer of CodeBERT. Each subtoken is mapped to a fixed-size vector $( \mathbf{e}_i \in \mathbb{R}^d ) (( d = 768 )) $, to reduce memory and computation overhead, a linear projection layer is applied:

\begin{equation}
    \mathbf{e}_i' = W_{\text{proj}} \cdot \mathbf{e}_i + b_{\text{proj}}
\end{equation}
where \( W_{\text{proj}} \in \mathbb{R}^{256 \times 768} \) is the projection matrix, and \( b_{\text{proj}} \) is the bias term. The resulting vector \( \mathbf{e}_i' \in \mathbb{R}^{256} \) serves as the final input representation for each subtoken. 

These input embeddings are passed through the transformer encoder layers of CodeBERT, where each layer applies a multi-head self-attention mechanism to compute the relationships between different tokens in the sequence. 
L2 regularization is applied to all learnable parameters to prevent overfitting during fine-tuning. 
Additionally, a dropout layer with a rate of 0.1 is applied after the feed-forward neural network (FFN) layers to prevent overfitting further. The FFN computation is given by:
\begin{equation}
    \text{FFN}(x) = \text{ReLU}(W_1x + b_1)W_2 + b_2
\end{equation}
where \( W_1 \) and \( W_2 \) are weight matrices, and \( b_1 \) and \( b_2 \) are bias vectors.
After passing through multiple encoder layers, CodeBERT produces an embedding matrix \( \mathbf{H} \) for the entire sequence of nodes:
\(\mathbf{H} = [\mathbf{h}_1, \mathbf{h}_2, \ldots, \mathbf{h}_N]\) 
where \( \mathbf{h}_i \in \mathbb{R}^d \) represents the feature vector for the \( i \)-th AST node, and \( N \) is the number of nodes in the AST.
To obtain a global feature vector representing the entire AST, a mean pooling operation is applied over all node-level embeddings:
\[
\mathbf{x}_{\text{ast}} = \frac{1}{N} \sum_{i=1}^{N} \mathbf{h}_i
\]
This pooling operation aggregates the individual node embeddings into a single vector \( \mathbf{x}_{\text{ast}} \) that captures the overall syntactic structure and variable dependencies of the smart contract.

\begin{figure}[ht]
    \centering
    \vspace{-5pt}
     \includegraphics[height=0.22\textheight]{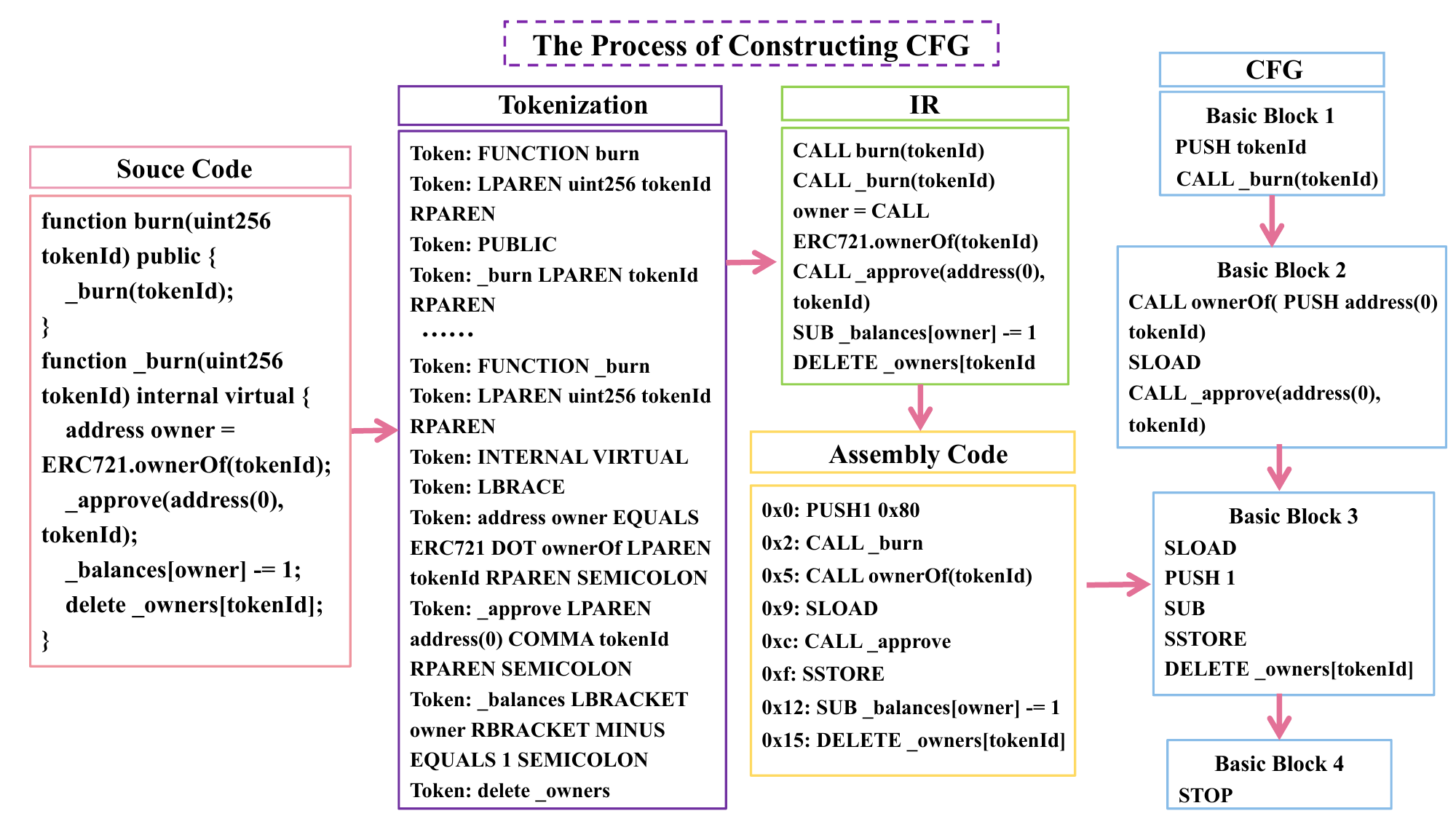}
    \caption{The Process of Constructing CFG}
    \vspace{-6pt}
    \label{fig2}
\end{figure}

However, relying solely on the AST feature vector is insufficient to fully capture the contract's control flow, cross-function interactions, and execution sequence. Therefore, in addition to extracting AST features, we also extract features from the Control Flow Graph (CFG). The CFG captures the execution flow within the program, showing the control dependencies between different code blocks, particularly in function calls, conditional branches, loops, and other control structures.
Each node in the CFG represents a basic block composed of multiple EVM instructions, with edges indicating the control flow paths between these basic blocks, summarizing the program's execution paths. The CFG visualizes the execution paths of the program and captures control dependencies, especially for function calls, conditional branches, and loops. It is represented as a directed graph \( G=(V, E) \), where \( V \) represents the nodes corresponding to basic blocks, each containing a set of sequential EVM instructions, and \( E \) represents the edges between nodes, indicating the control flow paths from one basic block to another. Figure \ref{fig2} illustrates the process of constructing the CFG.
To extract features for each basic block, we use a TextCNN\cite{textcnn2023svscanner} model to process the sequence of EVM instructions embedded using Word2Vec. Each basic block consists of a sequence of EVM instructions, represented as \( B_i = [I_{i1}, I_{i2}, \ldots, I_{iN}] \in \mathbb{R}^{N \times d} \), where \( N \) is the number of EVM instructions, and \( d \) is the embedding dimension. These instructions are embedded using a pre-trained Word2Vec model that maps each instruction to a fixed-length vector.
The TextCNN model applies K convolutional filters, each with $R$ convolutional kernels of varying sizes \( h \times d \) to capture features at different scales. Formally, the output feature vector obtained by the \( k \)-th filter is represented as:
\begin{equation}
     f_{i}^{(k)} = \text{ReLU}(W_k \ast B_i + b_k)
    \label{eq4}
\end{equation}
where \( W_k \in \mathbb{R}^{h \times d} \) is the weight matrix of the \( k \)-th convolutional filter, \( b_k \) is the bias, and \( f_{i}^{(k)} \) is the feature vector after applying the activation function ReLU. Each filter slides over the input sequence to aggregate local structural features of varying lengths.
After obtaining the feature maps, a max-pooling operation is performed to extract the most significant feature from each feature map:
\(h_{\text{block}} = \max(f_{i}^{(1)}, f_{i}^{(2)}, \ldots, f_{i}^{(M)})\) 
where \( h_{\text{block}} \) is the global feature vector for the basic block. The feature vectors for all basic blocks are concatenated to form the complete feature matrix \( C \).The loss function used is the cross-entropy loss, defined as:

\begin{equation}
    \mathcal{L}_{\text{CE}} = -\sum_{i=1}^{N} y_i \log(\hat{y}_i)
    \label{eq4}
\end{equation}
where \( y_i \) is the true label, and \( \hat{y}_i \) is the predicted probability for the $i-th$ sample.
This configuration ensures TextCNN effectively captures multi-scale local structural features from the EVM instruction sequences. 

After extracting these basic block features, the Graph Convolutional Network (GCN) \cite{gcn2021smart} processes the entire CFG using these features as node-level inputs to generate a global CFG representation. The feature vector of each node (basic block) $h_{\text{block}} $ is used as input to the GCN, represented as $H = \{ h_1, h_2, \ldots, h_n \}$. 
Each edge $e_k$ in the CFG is embedded as a vector $e_k$, which encodes the relationship between nodes, such as function calls, conditional branches, and loops. To enable the transmission of information between nodes, the feature vector of the start node $h_{\text{start}_k}$ is concatenated with the edge embedding $e_k$ to form the input message vector $x_k$:
\( x_k = [h_{\text{start}_k} || e_k] \)
where $ ||$ denotes the concatenation operation. This step fuses the node features and edge information, allowing the message passing process to capture contextual control flow information.
Next, the GCN computes the message $m_k$ transmitted along each edge using a message generation network:
\begin{equation}
    m_k = W_m'\cdot x_k + b_m'
    \label{eq4}
\end{equation}
where $W_m'$  is a learnable weight matrix, and  $b_m'$ is a bias term. The message $m_k$ represents the information passed from the start node to the end node of the edge.

To update the feature representation of each node $ h_v^{(l+1)}$, the aggregated messages $\sum_{u \in \mathcal{N}(v)} m_u$ from its neighboring nodes are combined with the node's features $h_v^{(l)} $ using a non-linear transformation:
\begin{equation}
    h_v^{(l+1)} = \phi \left( W_v' \cdot \sum_{u \in \mathcal{N}(v)} m_u + U_v' \cdot h_v^{(l)} + b_v' \right)
    \label{eq2}
\end{equation}
where $\phi $ is the ReLU activation function, and  $W_v'$, $ U_v'$, and $b_v'$ are learnable parameters.  Integration of information from neighboring nodes and the updating of the node's state.  
During the node update process, to distinguish the impact of different neighboring nodes, the GCN introduces a multi-head attention mechanism that assigns varying weights to neighboring nodes during the feature aggregation step:
\begin{equation}
    \beta_{vu} = \frac{\exp \left( \text{LeakyReLU}(q^T [P h_v' || P h_u']) \right)}{\sum_{w \in \mathcal{N}(v)} \exp \left( \text{LeakyReLU}(q^T [P h_v' || P h_w']) \right)}
    \label{eq3}
\end{equation}
where $\beta_{vu}$ represents the attention weight assigned to the neighbor node $ u$ for the target node $v $, and $ q $and  $P$ are learnable parameters.
Once all node features are updated, the GCN applies a weighted pooling operation to generate a global feature representation of the control flow graph (CFG):
\begin{equation}
    X_{\text{CFG}} = \sum_{v \in V} \beta_v h_v^{(L)}
    \label{eq4}
\end{equation}
where $\beta_v $ is the attention weight for node $v$ , indicating its contribution to the global graph embedding. This pooling operation adaptively aggregates the node-level features, producing a comprehensive global embedding of the CFG that effectively captures control flow semantics.

\begin{figure}[!ht]
    \centering
    \includegraphics[width=1\linewidth]{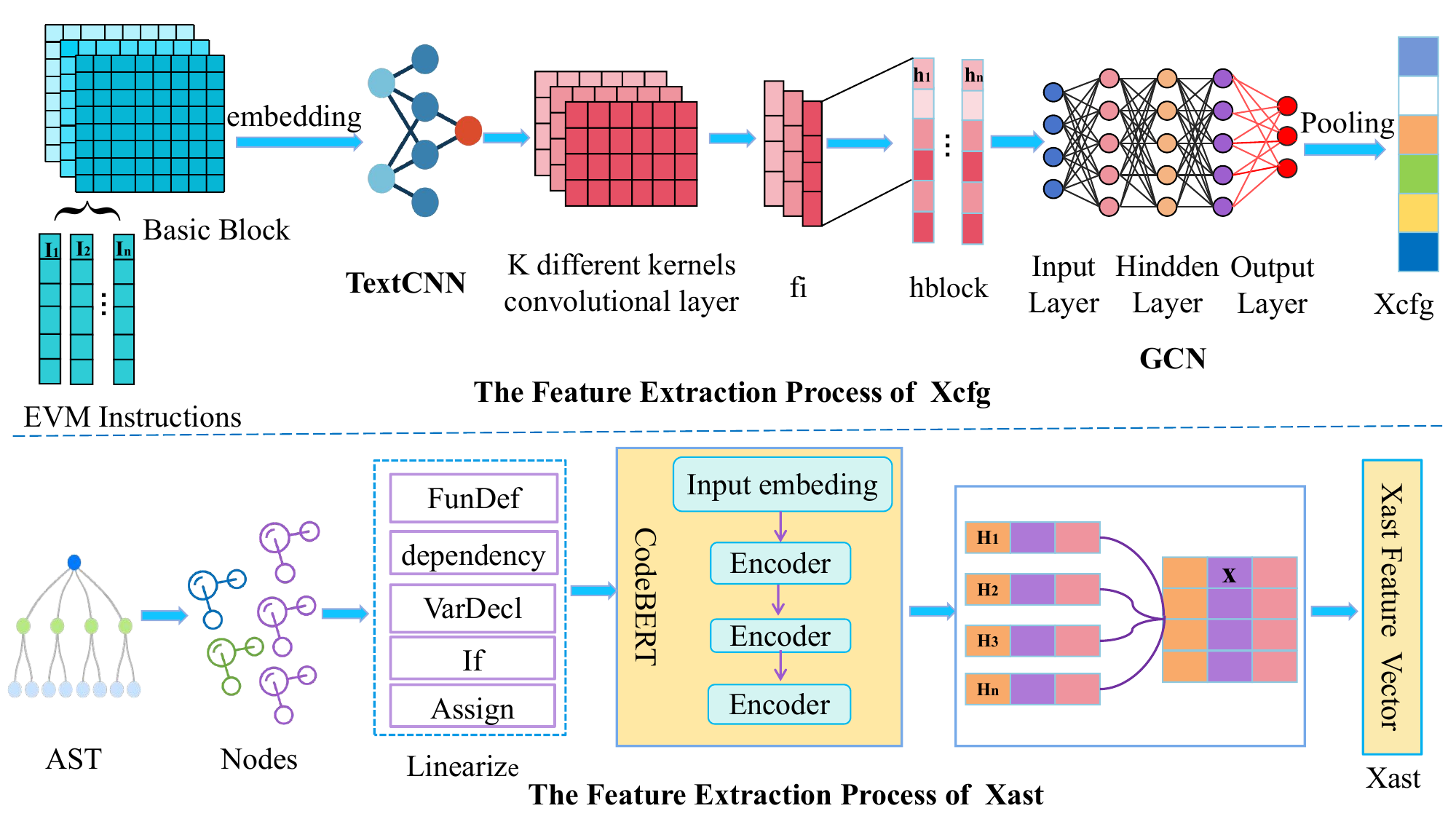}
    \caption{The Feature Extraction Process of $X_{ast}$ \& $X_ {cfg}$}
    \vspace{-6pt}
    \label{fig3}
\end{figure}

To ensure compatibility during feature fusion, a linear transformation is applied to the global feature vectors of both the AST and CFG. Specifically, the AST feature vector $X_{\text{AST}}$and the CFG feature vector $X_{\text{CFG}}$  are mapped to a common dimensional space.
$W_{\text{AST}} \in \mathbb{R}^{d' \times d_{\text{AST}}} $ and $W_{\text{CFG}}\in \mathbb{R}^{d' \times d_{\text{CFG}}}$ are learnable weight matrices, and $b_{\text{AST}},b_{\text{CFG}} $ are bias terms. This transformation aligns the dimensions of the AST and CFG representations, ensuring that they are compatible with subsequent combinations.
After the linear transformation, the normalized feature vectors are fused to form a comprehensive feature representation:
\begin{equation}
    X_{\text{combined}} = \alpha \cdot X_{\text{AST}}' + (1 - \alpha) \cdot X_{\text{CFG}}'
    \label{eq6}
\end{equation}
where $\alpha $ is a learnable parameter that controls the relative contribution of the AST and CFG features. 
The model is trained to optimize the feature fusion weights by minimizing the loss function:
\begin{equation}
     \min_{W_{cfg}, W_{ast}} \sum_{i=1}^{N} \mathcal{L} \left( f(\mathbf{x}_{\text{combined}}^{(i)}), y^{(i)} \right)
    \label{eq6}
\end{equation}
where \( \mathcal{L} \) denotes the loss function (e.g., cross-entropy), \( f(\mathbf{x}_{\text{combined}}^{(i)}) \) represents the model's predicted output for the \( i \)-th sample, and \( y^{(i)} \) is the corresponding true label.

After obtaining the optimized combined feature vector $\mathbf{x}_{combined}$, the next step involves integrating it with known defect information. Known defect code snippets are converted into high-dimensional vector representations using a pre-trained text embedding model (text-embedding-004). This model processes each defect code snippet and outputs an embedding vector $\mathbf{d}_{embedding}$, that encodes semantic information related to the defect type.
To address the issue of increased feature dimensionality when concatenating $\mathbf{x}_{combined}$ and $\mathbf{d}_{embedding}$, we introduce a dimension alignment module before the concatenation step to project both feature vectors into a unified low-dimensional space:

\begin{equation}
    \mathbf{x}_{\text{combined}}' = W_{\text{low}}^{(\text{combined})} \cdot \mathbf{x}_{\text{combined}} + b_{\text{low}}^{(\text{combined})}
    \label{eq6}
\end{equation}

\begin{equation}
    \mathbf{d}_{\text{embedding}}' = W_{\text{low}}^{(\text{embedding})} \cdot \mathbf{d}_{\text{embedding}} + b_{\text{low}}^{(\text{embedding})}
    \label{eq6}
\end{equation}
where \( W_{\text{low}}^{(\text{combined})} \) and \( W_{\text{low}}^{(\text{embedding})} \) are learnable projection matrices, and \( b_{\text{low}}^{(\text{combined})} \), \( b_{\text{low}}^{(\text{embedding})} \) are bias terms. 
The attention-weighted combination is performed as follows:
\begin{equation}
    \mathbf{E} = \alpha \cdot \mathbf{x}_{\text{combined}}' + (1 - \alpha) \cdot \mathbf{d}_{\text{embedding}}'
    \label{eq6}
\end{equation}
where \( \alpha \) is a learnable parameter the structural features from $\mathbf{x}_{\text{combined}}' $and the semantic features from $\mathbf{d}_{\text{embedding}}'$.

The resulting matrix  $E$ is constructed as follows:
\(\mathbf{E} = \left[ \mathbf{E}_1, \mathbf{E}_2, \ldots, \mathbf{E}_k \right]\)
where \( \mathbf{E}_k \) represents the combined feature embedding for the \( k \)-th known defect type. 
Each defect type has its own corresponding defect embedding, ensuring that the representation in $E$ is comprehensive and covers various vulnerabilities. These combined matrices, including matrix $E$, are stored in the vector database (VecDB). 
Table \ref{table1} describes the definitions and corresponding scenarios for three common NFT smart contract defects.

\subsection{Knowledge Base Retrieval}
 \begin{figure}[h!]
    \centering
     \vspace{-6pt}
    \includegraphics[width=1\linewidth]{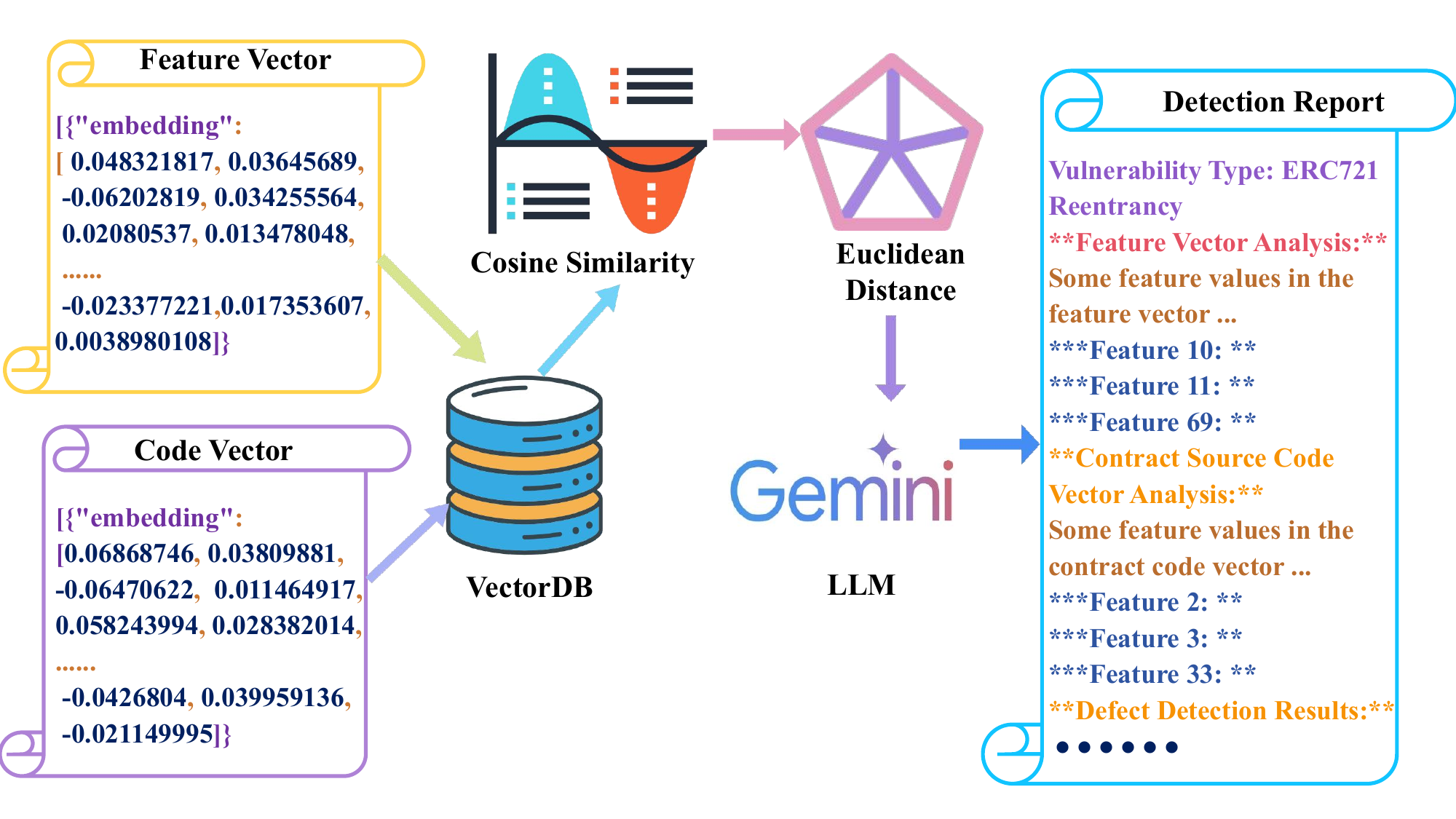}
      \vspace{-6pt}
    \caption{LLM Reasoning Process Detection Report }
      \vspace{-6pt}
    \label{fig4}
\end{figure}
During the knowledge base retrieval process, the system compares the generated vectors with the known defect vectors stored in the VectorDB. To evaluate the similarity between the vectors, the tool combines the use of cosine similarity and Euclidean distance. Cosine similarity effectively measures the similarity between vectors by calculating the angle between them, with values closer to 1 indicating higher similarity. Euclidean distance, on the other hand, considers the absolute magnitude of the vectors, capturing quantitative differences in contract operations such as the number of minted tokens or function call counts. 
The following is the mathematical formula for cosine similarity. Cosine similarity ($\cos \theta $) measures the similarity between two vectors by calculating the dot product of the vectors divided by the product of their magnitudes. The formula is given as follows:
\begin{equation}
    \cos \theta = \frac{\mathbf{A} \cdot \mathbf{B}}{\|\mathbf{A}\| \|\mathbf{B}\|} = \frac{\sum_{i=1}^{n} A_i B_i}{\sqrt{\sum_{i=1}^{n} A_i^2} \sqrt{\sum_{i=1}^{n} B_i^2}}
    \label{eq7}
\end{equation}
where:$\mathbf{A}$ and $\mathbf{B}$ represent the contract vector under analysis and the known defect vector, $A_i$ and $B_i$ are the $i$-th components of vectors $\mathbf{A}$ and $\mathbf{B}$, respectively, $n$ is the dimension of the vectors.
To capture the absolute magnitude of feature values, we introduce Euclidean distance, which is calculated as follows:
\begin{equation}
    d\left (  A,B \right ) =\sqrt{\sum_{i=1}^{n}\left ( A_{i} -B_{i} \right )^{2} } 
    \label{eq8}
\end{equation}

This formula measures the absolute distance between two vectors in feature space, allowing us to consider the structural similarity of the contracts and their differences in quantity and scale.
As illustrated in Figure \ref{fig4}, the process begins by selecting defect types with high similarity from the knowledge base retrieval phase, where cosine similarity is calculated to measure the alignment between the contract vectors and known defect vectors. Euclidean distance is employed to capture the magnitude differences between the contract and defect vectors.

\renewcommand{\arraystretch}{1}
\begin{table}[ht]
\centering
\vspace{-6pt}
\caption{The definitions for NFT smart contract defects}
\footnotesize
\begin{tabular}{p{1.6cm} p{2.2cm} p{4.5cm}}
\toprule
\textbf{Defect Type} & \textbf{Definition} & \textbf{Scenario Description} \\
\midrule
ERC-721 Reentrancy  & Modifies the state variable after the external invocation.\cite{yang2023definition} & A malicious onERC721Received function can reenter the victim contract during a token transfer, disrupting its logic. This may allow the minting of more NFTs than intended, causing economic loss.\\

Public Burn & Does not check the caller of the burn operation on NFT.\cite{yang2023definition} & If the burn function is not restricted to the NFT owner, anyone could burn another's NFT without permission. This could lead to the unintended destruction of all NFTs in a project.\\

Risky Mutable Proxy & Makes the proxy contract modifiable.\cite{yang2023definition} & If the proxy registry address is modifiable, an attacker could change it and transfer all tokens without permission. This poses a significant risk as the proxy can act on behalf of the user.\\

Unlimited Minting & When there is no check on the max supply during the minting process.\cite{yang2023definition} & Without verifying whether the current supply exceeds the promised limit, the contract owner or an attacker could exploit a reserve function to mint unlimited NFTs.\\

\bottomrule
\end{tabular}%
\label{table1}%
\end{table}

 \vspace{-3.8pt}
\subsection{LLM Reasoning}

In the LLM reasoning phase, NATLM performs the final stage of defect detection and analysis using a large language model (LLM). 
Once the most relevant defect types are identified through the combination of cosine similarity and Euclidean distance, these feature vectors, along with the contract’s feature and code vectors, are input into the LLM for in-depth analysis. The LLM further interprets and validates the similarity, deeply analyzing the contextual relationships between the identified defect type and the current contract.
The similarity-based retrieval results by adjusting the model’s sensitivity to imbalanced data distributions through a weighted loss function. The weight assigned to each defect type is calculated as:  

\begin{equation}
    w_i = \frac{1}{\log(1 + n_i)}
    \label{eq10}
\end{equation}
where \( n_i \) represents the sample count for the \( i \)-th defect type. This weighting scheme ensures that defect types with fewer samples receive higher weights, thereby increasing their influence during gradient updates and improving the model’s performance on rare defect types.
During the training process, the weighted loss function is defined as:
\begin{equation}
    \mathcal{L}_{\text{weighted}} = - \sum_{i=1}^{N} w_i y_i \log(\hat{y}_i)
\end{equation}
where \( N \) is the total number of samples, \( y_i \) and \( \hat{y}_i \) represent the true label and predicted probability for the \( i \)-th sample, respectively, and \( w_i \) is the corresponding weight for the defect type. 
The feature vectors of these identified defects, along with the feature vectors and code vectors of the current contract, are then input into the LLM for in-depth reasoning.
To improve the accuracy of defect detection results and reduce false positives, a confidence threshold filtering strategy is introduced during the output stage. To prevent low-confidence predictions from affecting the analysis report, an adjustable confidence threshold \( \tau \) is set. When the confidence score \( \hat{p}_i \) of a defect type exceeds \( \tau \), the result is retained and recorded in the analysis report. Conversely, predictions with confidence scores below \( \tau \) are filtered out to prevent unreliable results. This filtering process follows the rule:
\begin{equation}
    \hat{y}_i = 
     \begin{cases}
       y_i, & \text{if } \hat{p}_i \geq \tau, \\
       \varnothing, & \text{if } \hat{p}_i < \tau,
\end{cases}
    \label{eq25}
\end{equation}
where \( \hat{p}_i \) denotes the confidence score for the \( i \)-th defect type, and \( y_i \) represents the corresponding predicted label.
After filtering low-confidence predictions, the LLM generates a comprehensive detection report. This report includes a detailed analysis of the identified defect types, their potential security impacts, and recommended remediation measures. 
 \vspace{-2pt}
\section{Experiment}

To construct a dataset of NFT smart contracts, we utilized the publicly available Smart Contract Sanctuary on GitHub \cite{smart_contract_sanctuary}, which aggregates verified smart contracts from multiple blockchain networks (e.g., Ethereum, BSC, Polygon, etc.). The contracts are organized by their unique addresses and grouped by network. To extract relevant NFT smart contracts, we apply a pattern-matching approach based on ERC-721 interface definitions and related keywords. Specifically, we search for interface function signatures, such as supportsInterface(bytes4 interfaceID) and ownerOf(uint256 tokenId), alongside keywords such as "NFT" and "ERC721" in contract names, comments, and function signatures. We compared extracted contracts against interface definitions to confirm adherence to relevant standards.
We incorporate data from \cite{yang2023definition} to the dataset. To avoid redundancy when merging the two datasets, we compare the addresses of the contracts and remove duplicates. After applying consistent filtering and deduplication criteria, we obtain a dataset containing 8,672 NFT smart contracts.
In the experimental setup, CodeBERT adopts a two-layer multi-layer perceptron (MLP) and is trained for 30 epochs with a batch size of 128, a learning rate of 0.0003, a dropout rate of 0.1, and the AdamW optimizer. TextCNN is used to extract initial features from the basic blocks in the CFG, utilizing the Adam optimizer, with a batch size of 256, and a learning rate of 0.0003. The GCN, which processes the CFG to capture relational and structural information, consists of two convolutional layers, each with a hidden dimension of 128. The GCN is trained using the Adam optimizer, with a batch size of 64, a learning rate of 0.001, and a dropout rate of 0.5.
\begin{table*}[h]
\centering
  \vspace{-6pt}
\caption{Comparison of Detection Capabilities of NATLM and Baseline Tools Smart Contract Vulnerabilities. \\
\textbf{\ding{51}} indicates always identifies \textbf{O} indicates partial identification, and \textbf{\ding{55}} indicates low precision or complete failure to detect this vulnerability.}
\renewcommand{\arraystretch}{1.18} 
\setlength{\tabcolsep}{10pt} 
\resizebox{\linewidth}{!}{
\begin{tabular}{l|c|c|c|c|c|c}
\hline
\textbf{Tool} & \textbf{Reentrancy} & \textbf{Access Control} & \textbf{ERC-721 Reentrancy} & \textbf{Public Burn} & \textbf{Risky Mutable Proxy} & \textbf{Unlimited Minting} \\ 
\hline
Mythril & O & O & \ding{55} & \ding{55} & \ding{55} & \ding{55} \\
Securify & \ding{51} & \ding{55} & \ding{55} & \ding{55} & \ding{55} & \ding{55} \\
Oyente & \ding{51} & \ding{55} & \ding{55} & \ding{55} & \ding{55} & \ding{55} \\
Slither & \ding{51} & \ding{55} & \ding{55} & \ding{55} & \ding{55} & \ding{55} \\
Sailfish & \ding{51} & \ding{55} & \ding{55} & \ding{55} & \ding{55} & \ding{55} \\
AChecker & \ding{55} & \ding{51} & \ding{55} & \ding{55} & \ding{55} & \ding{55} \\
GPT-3.5-turbo & O & \ding{55} & \ding{55} & \ding{55} & \ding{55} & \ding{55} \\
GPT-4 & O & O & O & O & O & O \\
GPT-4o & O & O & O & O & O & O \\
Gemini pro-1.5 & O & O & O & O & O & O \\
\bfseries NATLM  & \ding{51} & \ding{51} & \ding{51} & \ding{51} & \ding{51} & \ding{51} \\
\hline
\end{tabular}
}
 \vspace{-6pt}
\label{table 2}
\end{table*}

\begin{table*}[h!]
\centering
\caption{The performance metrics of NATLM are compared against four LLM models, evaluated based on  Precision (PRE), Recall (RE), and F1-score (F1).}
\resizebox{\linewidth}{!}{
\renewcommand{\arraystretch}{1.88}
\begin{tabular}{c|ccc|ccc|ccc|ccc|ccc}
\hline
\multicolumn{1}{c|}{} & \multicolumn{3}{c|}{ Reentrancy} 
& \multicolumn{3}{c|}{ERC-721 Reentrancy} & \multicolumn{3}{c|}{Public Burn} & \multicolumn{3}{c|}{Risky Mutable Proxy} & \multicolumn{3}{c}{Unlimited Minting}  \\ \cline{2-16}

Tool & PRE & RE & F1 & PRE & RE & F1 & PRE & RE & F1 & PRE & RE & F1  & PRE & RE & F1\\ \hline
GPT-3.5-turbo & 26.03\% & 38\% &  30.89\% & 23.5\% & 41.96\% &  30.12\% & 19.16\% & 35\% &  24.76\% & 15.2\% & 28.51\% & 19.83\% & 17.5\% & 32.61\% &  22.78\% \\ 

GPT-4 & 33.5\% & 87.1\% &  48.39\% & 32.02\% & 82.8\% &  46.18\% & 25.12\% & 80.2\% & 38.26 \% & 28.91\% & 78.75\% & 42.29\% & 30.51\% & 83.55\% &  	44.7\% \\ 

GPT-4o & 30.17\% & 83.92\% &   44.38\% & 28.2\% & 79.61\% &  41.65\% & 22.5\% & 77.43\% &  34.86\% & 30.5\% & 74.8\% & 43.32\% & 33.19\% & 80.92\% &  47.07\% \\ 

Gemini pro-1.5 & 31.76\% & 80.2\% &  45.5\% & 31.62\% & 81.5\% &  45.56\% & 32.18\% &76.5\% &  45.3\% & 32.89\% & 75.08\% &  45.74\% & 34.17\% & 82.52\% &  48.33\% \\ 

\hline
\bfseries NATLM &\bf 87.12\% &\bf 90.58\% &\bf 88.82\% & \bf 87.75\% &\bf 84.09\% & \bf 85.88\% & \bf 85.74\% & \bf 95.45\% &\bf 90.32\% & \bf 92.85\% &\bf 86.66\% &\bf 89.65\% &\bf 85.16\% &\bf 91.16\% &\bf 88.06\%\\ 
\hline
\end{tabular}
}
 \vspace{-6pt}
\label{table 3}

\end{table*}

 \begin{figure}[h!]
    \centering
    \vspace{-6pt}
    \includegraphics[width=1\linewidth]{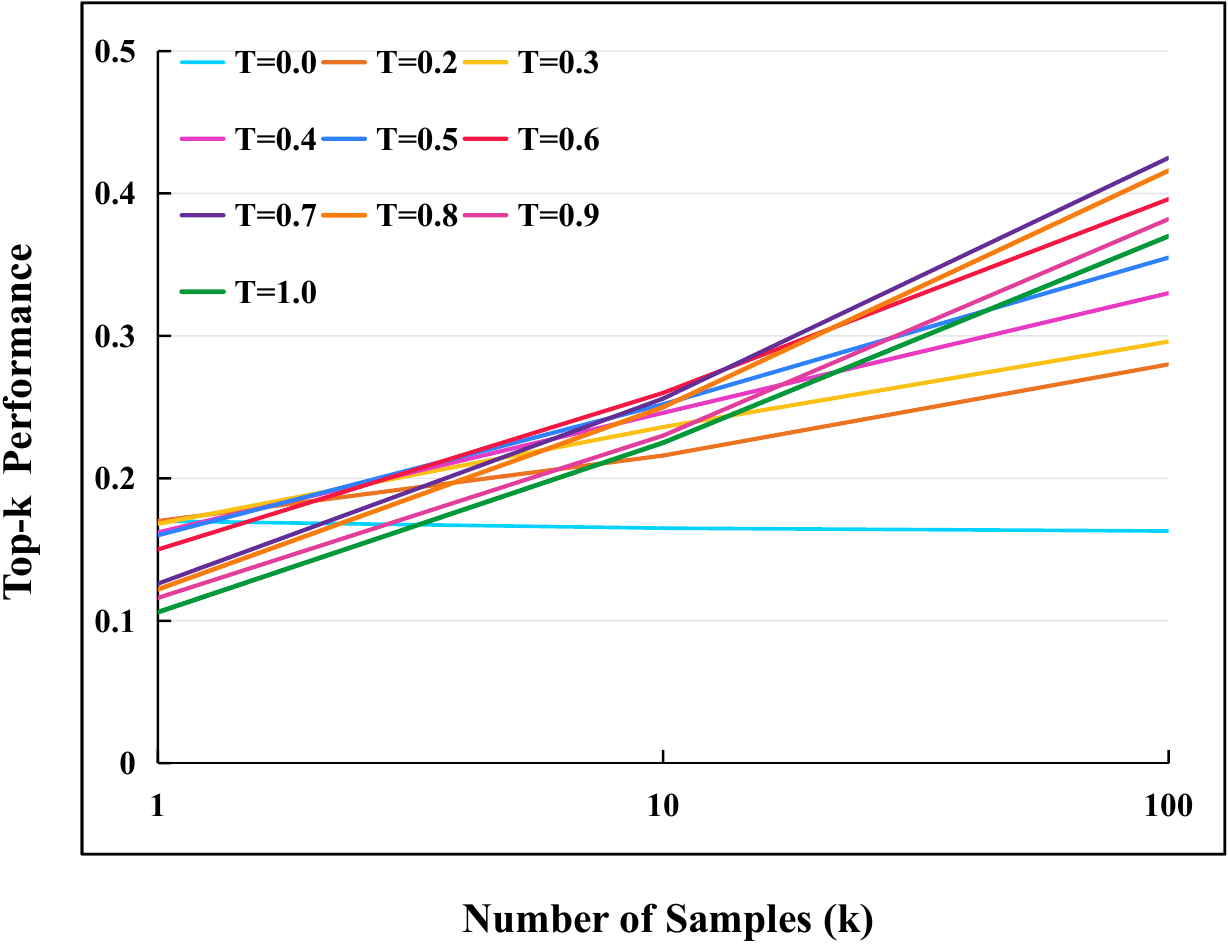}
    \caption{Effect of Temperature on Top-K Performance Across Samples}
      \vspace{-8pt}
    \label{fig5}
\end{figure}

Temperature is an important hyperparameter that controls the randomness of model-generated results. When the temperature is high, the diversity of the generated results increases, but this may come at the cost of accuracy. Conversely, when the temperature is low, the generated results tend to be consistent with reduced randomness. As shown in Figure \ref{fig5}, the chart presents the model's performance under different temperature (T) settings while generating different samples. As the number of generated samples (k) increases, a higher temperature setting (T=0.7) achieves the best Top-k Performance, indicating that in scenarios with a large number of samples, a higher temperature helps improve the quality of the generated audit reports.

To compare the detection performance of NATLM with state-of-the-art (SOTA) smart contract vulnerability detection tools, we selected 10 vulnerability detection tools. These tools include symbolic execution-based tools (Mythril, Oyente\cite{Oyente}), static analysis tools ( Securify\cite{2018securify}, Slither, Sailfish\cite{2022sailfish}, AChecker\cite{2023achecker}), and machine learning-enhanced LLMs for smart contract analysis, specifically targeting NFT-related vulnerabilities. The comparison also evaluates the detection capabilities of LLMs models (GPT-3.5-turbo, GPT-4, GPT-4o, and Gemini pro-1.5) for NFT-specific vulnerabilities. Some tools are constrained to predefined vulnerability categories and cannot generalize to detect vulnerabilities beyond their predefined rules. In contrast, LLM tools demonstrate the potential to adapt to new vulnerability categories.
Table \ref{table 2} illustrates the performance of NATLM and 10 baseline detection tools across six smart contract vulnerability categories: Reentrancy, Access Control, ERC-721 Reentrancy, Public Burn, Risky Mutable Proxy, and Unlimited Minting. The results indicate that while traditional symbolic execution and static analysis tools perform well for conventional vulnerabilities (such as reentrancy), they struggle to detect NFT-specific vulnerabilities (e.g., ERC-721 reentrancy and unlimited minting).
In contrast, LLMs demonstrate greater adaptability across different categories. However, they still suffer from moderate rates of false positives and negatives. NATLM outperforms traditional tools and LLM baselines by consistently achieving high precision and recall across all types of vulnerabilities.

In the experiment, we categorize the detection results of NATLM into three classes: True Positive (TP), False Positive (FP), and False Negative (FN). The results, there are 503 NFT smart contracts with ERC-721 Reentrancy defects, 44 with Public Burn defects, 15 with Risky Mutable Proxy defects, and 781 with Unlimited Minting defects in the dataset. NATLM detects 482 ERC-721 Reentrancy defects, of which 423 are correctly classified as true positives (TP), 59 as false positives (FP), and 80 as false negatives (FN). For the Public Burn defect, NATLM detects 49 contracts with the defect, with 42 correctly classified as TP, 7 as FP, and 2 as FN. For the Risky Mutable Proxy defect, NATLM detects 14 contracts with the defect, with 13 correctly classified as TP, 1 as FP, and 2 as FN. Finally, for the Unlimited Minting defect, NATLM detects 836 contracts with the defect, with 712 correctly classified as TP, 124 as FP, and 69 as FN.

Table \ref{table 3} presents the comparative performance of NATLM and four LLM baselines(GPT-3.5-turbo, GPT-4, GPT-4o, and Gemini pro-1.5) across five categories of smart contract vulnerabilities: Reentrancy, ERC-721 Reentrancy, Public Burn, Risky Mutable Proxy, and Unlimited Minting. The evaluation is conducted on a dataset comprising 8,672 NFT smart contracts.
For Reentrancy, GPT-3.5-turbo performs poorly, with an F1-score of 30.89\%, precision of 26.03\%, and recall of 38\%. This suggests that GPT-3.5-turbo may struggle to distinguish between vulnerable and non-vulnerable contracts, potentially due to limitations in its contextual understanding or token length constraints. In contrast, GPT-4 shows a substantial improvement in recall (87.1\%), but its precision remains low at 33.5\%. This indicates that while GPT-4 can detect the most vulnerable contracts, it generates many false positives.
For ERC-721 Reentrancy, NATLM achieves an F1-score of 85.88\%, significantly outperforming GPT-4 (46.18\%) and Gemini pro-1.5 (45.56\%). The relatively low precision scores of the LLM baselines (ranging from 23.5\% to 32.02\%). 
NATLM achieves an F1-score of 90.32\%  for Public Burn vulnerabilities, indicating a highly accurate detection of unauthorized token destruction. In comparison, GPT-4 achieves a recall of 80.2\% but a much lower precision (25.12\%), resulting in an F1-score of 38.25\%. This suggests that GPT-4 can detect many true vulnerabilities, with a high false positive rate.
For Risky Mutable Proxy and Unlimited Minting, NATLM maintains F1-scores of 89.65\% and 88.06\%. In contrast, the LLM baselines achieve F1-scores ranging from 32\% to 48\%, with GPT-4o demonstrating slightly better recall but consistently low precision.
Overall, NATLM outperforms the baseline tools in detecting vulnerabilities in NFT smart contracts, achieving an overall Precision of 87.72\%. 

\section{Reproducibility}
 To ensure the reproducibility of our study, we provide all necessary resources, including:(1)All datasets drawn from the existing literature (potentially including authors' own previously published work) are publicly available. (2)Full details of the experimental setup as outlined in Section 4. (3)All code required for conducting experiments will be made publicly available upon publication of the paper.

\section{ Conclusion}

In this study, we propose the NATLM framework, which combines static analysis and LLM (Gemini Pro 1.5) to detect defects in NFT smart contracts. The strength of the NATLM framework lies in its combination of structured code analysis from static analysis with the deep semantic understanding capabilities of the LLM. We conducted experiments on a dataset of 8,672 NFT smart contracts, achieving an overall precision of 87.72\%.

\bibliographystyle{named}
\bibliography{main}

\end{document}